\documentclass[11pt]{article}
\usepackage{epsf}
\usepackage{epsfig}  
\usepackage{url}

\input{psfig}
\newcommand{\mycomment}[1]{}
\newcommand{\myfig}[3]
{
\begin{figure}
\centerline{\psfig{figure=#1.eps,width=#2}}
\caption{#3}
\label{#1:fig}
\end{figure}
}

\begin{document}
\ifx\href\undefined\else\hypersetup{linktocpage=true}\fi 

\title{CDN: Content Distribution Network\thanks{Technical Report TR-125 of Experimental Computer Systems Lab in Stony Brook University.}}
\author{
Gang Peng \\ 
Department of Computer Science, \\
State University of New York at Stony Brook, \\
Stony Brook, NY 11794-4400. \\
{\tt gpeng@cs.sunysb.edu} }
\maketitle

\begin{abstract}
Internet evolves and operates largely without a central coordination,
the lack of which was and
is critically important to the rapid growth and evolution of Internet.
However, the
lack of management in turn makes it very difficult to guarantee
proper performance and to deal systematically with performance problems.
Meanwhile, the available network bandwidth and server capacity
continue to be overwhelmed by the skyrocketing Internet utilization and
the accelerating growth of bandwidth intensive content.
As a result, Internet service quality perceived 
by customers is largely unpredictable and unsatisfactory.
\emph{Content Distribution Network} (CDN) is an effective 
approach to improve Internet service quality.
CDN replicates the content from the
place of origin to the replica servers scattered over the Internet and serves
a request from a replica server close to where the request originates.
In this paper, we first give an overview about CDN.
We then present the critical issues involved in designing and
implementing an effective CDN and survey the approaches proposed in literature
to address these problems. An example of CDN is described to show how
a real commercial CDN operates. After this, we present a scheme that provides
fast service location for peer-to-peer systems,
a special type of CDN with no infrastructure support.
We conclude with a brief
projection about CDN.
\mycomment{We go on to propose a scheme to...}


\end{abstract}

\setcounter{page}{1}

\section {Introduction}

Over the past several years, the Internet has effectuated 
serious changes in how we transact business. For most businesses,
the performance
of Web connections has a direct impact on their profitability.
Although the time to
load the content of key Internet sites has improved constantly over the last
several years, overall Web content access latency is still in the range of
a few seconds, which is several times the
threshold believed to represent
natural human reading/scanning speeds. One might think that the constant
improvement in the bandwidth of Internet infrastructure, for example, the
availability of high-speed
"last mile" connection of the subscribers to the Internet and the backbone
fibers,
and the increasing capacity of the various servers would reduce or eliminate
the access
delay problem eventually.
However, the reality is quite the opposite. Even with these
improvements, users still suffer from very significant access delays.

The poor Internet service quality, typically represented as the long content
delivery delay, is primarily
due to two phenomena. The first is the lack
of overall management for Internet. Although the fact that the Internet evolves
and operates largely without central supervision was and is critically
important to the rapid growth and change,
the absence of overall administration in turn
makes it very difficult to guarantee proper performance and deal systematically
with performance problems. The second is that as the
load on Internet and the richness of its content continue to soar, any
increase in available network bandwidth and server capacity will
continue to be overwhelmed. These two facts not only elevate the delay in 
accessing content on Internet, but also make the access latency unpredictable.
With the emergence of new forms of Internet content, such as media
streams, the Internet performance problem will become even worse because
the fluctuation and irregularity of the access delay
will have a serious impact on the quality of the
streaming content that the customer perceives \cite{Gadde}.

The basic approach to address the performance problem is to
move the content from the
places of origin servers to the places at the edge of the Internet.
Serving content from a local replica server typically has better performance 
(lower access latency, higher transfer rate) than from
the origin server, and using multiple replica servers for servicing requests
costs less than using only the data communications network does.
CDN takes precisely this approach. In concrete, CDN
replicates a very selective set of content to the replica servers and
only sends those requests for the replicated content to a replica server.
\mycomment{Both Web caching system and CDN are the applications of using storage
to replicate content as a
means to accelerate Internet performance. They are complementary
in functionality to each other.
In Web caching system, the cache sees all requests from the
community of users served by the cache. The benefit from such a cache derives
from the commonality of Internet access within the
community -- the cache provides benefits if the requested object has already
been requested recently. In such an application, a cache demonstrates a
"hit ratio" on the order of 40\% \cite{CDD101, Abrams}. In contrast, CDN
replicates a very selective set of content to the replica servers and
only sends those requests for the replicated content to a replica server
so that all
requests hit. This higher hit ratio implies better performance of serving
requests compared to that using Web caching system solely.}
How to place replica servers and distribute content copies to replica servers
and how to route
the requests to the proper replica server having the desired content
are the key challenges in designing an
effective CDN, and are the major topics we will discuss.

The rest of the paper is organized as following: Section 2 gives an overview
about CDN, including the general CDN architecture and
the issues to be addressed in constructing a capable CDN.
In Section 3, we
describe the issues involved in replica placement in CDN and
compare some representative strategies.
We then proceed to Section 4, which discusses the problems we need to solve
for request routing and survey effective approaches proposed in literature to
resolve them.\mycomment{
as well as correlation between the content distribution and the
request routing.}
\mycomment{ Section 5 proposes a scheme... Section 6 Section 5 discusses
correlation between the content distribution and the request routing.}
Section 5 describes how a commercial CDN works and Section 6 presents a
request routing scheme for fast service location in
large-scale peer-to-peer systems.
We conclude in Section 6 with a brief projection about CDN.


\section{Overview}

CDN distributes the contents from the origin server to
the replica servers close to the end clients. The replica servers in a CDN
store a very selective set of content and only the requests for that
set of content are served by the CDN so that the hit ratio can approach 100\%.
This fact implies that CDN can present short access delay and consume
less network bandwidth. In addition,
CDNs offer compelling benefits to content providers,
including the popular Web sites \cite{Gadde}.
This is because a CDN can
serve multiple content providers, and the shared resources offer economies of
scale and allow the CDN to dynamically adjust content placement, request
routing and capacity provisioning to respond to demand and network conditions 
\cite{Gadde, CDD101, Day1}.  Moreover,
the fact that many objects are not cacheable but replicable,
which include dynamic objects with read-only access and personalized objects 
(e.g., "cookied" requests), makes CDN
indispensable.

\subsection{A General Architecture of CDN} 
\myfig{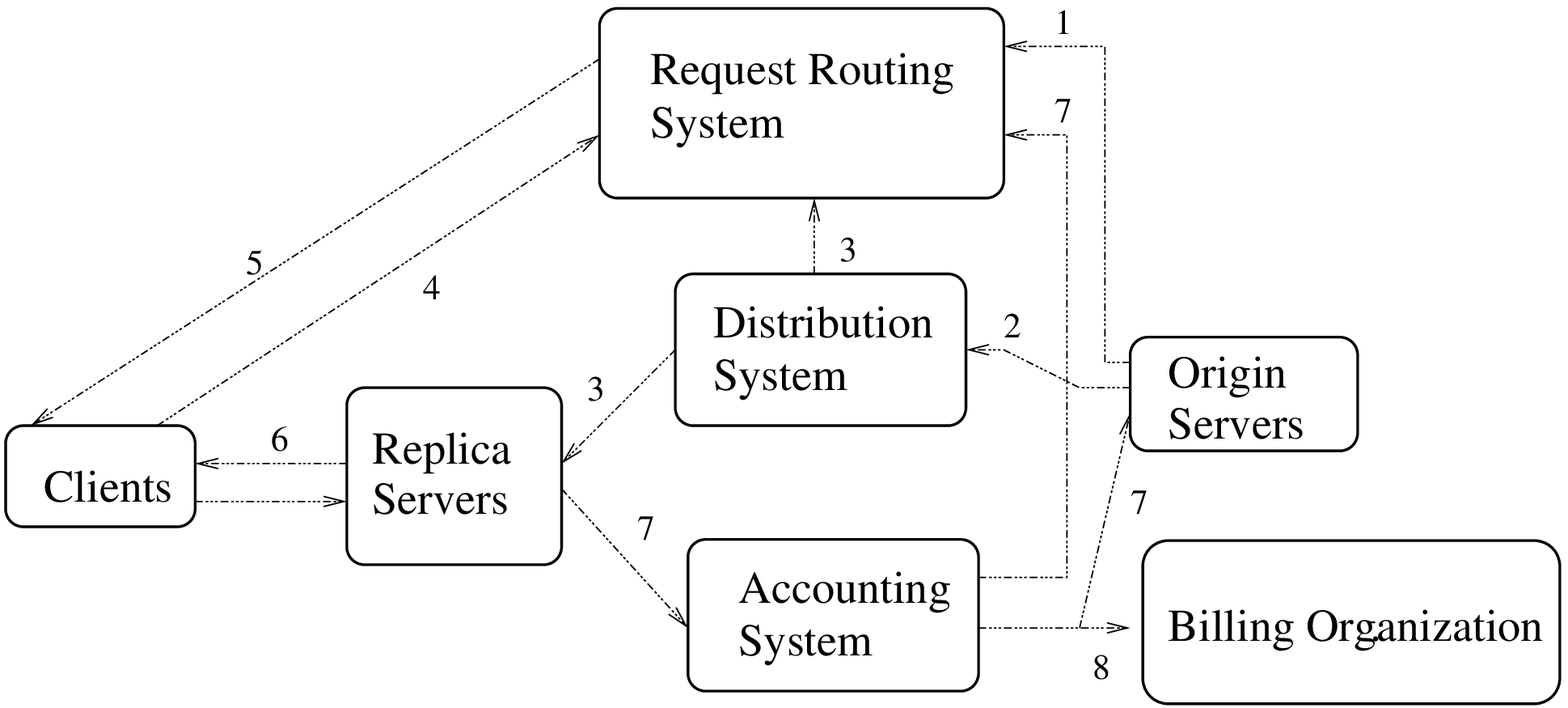}{4.2 in}{System Architecture Components of a CDN}
The general architecture of a CDN system is shown in Figure
\ref{cdn_arch:fig}. It
consists of seven components:
\emph{client}, \emph{replica servers},
\emph{origin server}, \emph{billing organization}, \emph{request routing system},
\emph{distribution system} and \emph{accounting system}. The relationships among
these components are represented with the numbered lines in Figure
\ref{cdn_arch:fig} and are described as
follows \cite{Green, Day1}:
\begin{enumerate}
\item The \emph{origin server}
delegates its URI name space for document objects to be
distributed and delivered by the CDN to the \emph{request routing system}.
\item The \emph{origin server} publishes content that is to be distributed and
delivered by the CDN into the \emph{distribution system}.
\item The \emph{distribution system} moves content to replica servers.
In addition, this system interacts with the \emph{request routing system}
through feedback to assist in the \emph{replica server} selection process for 
\emph{client} requests.
\item The \emph{client} requests documents from what it perceives to be the
origin. However, due to URI name space delegation, the request is actually
directed to the \emph{request routing system}.
\item The \emph{request routing system} routes the request to a suitable
\emph{replica server} in CDN.
\item The selected \emph{replica server} delivers the requested
content to the \emph{client}. Additionally, the \emph{replica server} sends
accounting information for delivered content to the \emph{accounting system}.
\item The \emph{accounting system} aggregates and distills the accounting
information into statistics and content detail records for use by the 
\emph{origin server} and \emph{billing organization}.
Statistics are also used as
feedback to the \emph{request routing system}.
\item The \emph{billing organization} uses the content detail records to settle
with each of the parties involved in the content distribution and delivery
process.
\end{enumerate}

\subsection{Distribution System}
There are two dominating approaches to distribute content to replica servers:
using the Internet, and using broadcast satellite. Internet distribution
of content is simpler. In such an approach, a CDN establishes and maintains a
distribution tree or an overlay network over the existing Internet
infrastructure
and disseminates content from the origin server to the replica servers via the
tree or overlay. How to establish and maintain the
distribution overlay or tree is the major
technical concern. This approach might suffer from the unpredictability and
problematic performance of the Internet itself. Akamai Technologies and
Sandpiper
Networks use this model. Data satellite broadcast has the potential for
remarkable cost savings, and it provides a high-quality, predictable performance
path for sending critical content such as real-time streaming media.
The content distribution schemes in CyberStar and Edgix utilize
data satellite broadcast.

\subsection{Replica Placement}
Replica placement deals with
how many replicas each object has and where in the
network to place them.
This problem can be further divided into two issues:
replica server placement and object replica placement.
Replica server placement deals with the problem of placing the replica servers
on the Internet. Object replica placement is about on which replica
server and how to place a particular object replica, e.g., a web page.
Intuitively, replica servers should be placed in such a manner that
they are \emph{closer} to the
clients, thereby reducing latency and bandwidth consumption. Also, object
replicas should be placed to even the load of the replica servers in CDN, that
is, trying to balance the load among replica servers \cite{Aggarwal}.

Some theoretical approaches are proposed to model the replica server placement
problem. These models are variations of or based on the
{\em center placement problem}. Due to the computational complexity of these
algorithms, heuristics have been developed. These suboptimal algorithms take
into account the existing information from CDN, such as the workload pattern
and the network topology. They provide sufficient solutions with less
computation cost.

Object replica placement has been well studied in Web caching system.
Korupolu et al. \cite{Korupolu1, Korupolu2} show that the cooperation between
caching nodes can improve the overall performance significantly.
For the object replica placement issue in CDN, Kangasharju et al. \cite{Kangasharju1} discussed a simple cost model and evaluated some heuristic approaches,
and proposed an improved heuristic exploiting the coordination between
replica servers. The essence of the work of Kangasharju et al.
is similar to the idea that Korupolu et al. present.

\mycomment{
More concretely, there are three issues in replica placement.
\begin{enumerate}
\item Which replica placement policy to choose, static replica placement or
dynamic replica placement? Static placement imposes less stringent requirements
for the request routing system. However, static placement is error prone, and
can become infeasible for large-scale system when the content get
changed frequently. On the other hand, dynamic replica
placement can adapt to large-scale system more easily.
\item What is the granularity of replication? Finer granularity imposes higher
overhead for collecting and maintaining access statistics and for placement
decision. Coarser granularity may increase the overhead for
disseminating objects to replica servers as. For example, if taking a site
as the unit to distribute content, some unpopular pages on a site may have to be
propagated to every replica server as the popular ones, which can be avoided
using finer content distribution granularity.
\item How to propagate replica placement information to the request routing
system in time? This problem is difficult to solve especially for the dynamic
placement policy because migration of objects among replica servers
may happen frequently and the database of storing the
location of replica objects used by the request routing system has to be kept
up-to-date accordingly.
\end{enumerate}
}

\subsection{Request Routing System} 
Request routing system is used to select a suitable replica server that
holds the
copy of the requested content and direct the incoming requests to that server.
Proximity between the client and the selected replica server and the
replica server
load are the two major criteria used to choose a proper
replica server.

\paragraph{Server location}
The question of how to choose a suitable replica server within a CDN to service
a given request involves the following issues:
\begin{itemize}
\item Determine the distance between the requesting client and a server. 
Hop counts and round-trip times are two often used metrics to measure
the distance. ``ping" and ``traceroute" are the common tools to obtain these two
parameters.
However, neither of these two metrics is sufficient and accurate to 
indicate the proximity between the clients and the replica servers because
the former does not account for the network traffic situation and the latter
is highly variable.
\item Determine the load of a replica server.
The techniques widely used to determine
the server load are {\em server push} and {\em client probe}.
In the first technique, the replica
servers propagate the load information to some agents. In the 
second approach, the agents probe the status of the servers of interest
periodically. There is trade-off between the frequencies of probing for
accurate measurement and the traffic incurred by probing.
\end{itemize}

\paragraph{Request routing}
Many techniques have been used to guide clients to use a particular
server among a set of replica servers. In general, they can be classified into
five categories.
\begin{itemize}

\item {\em Client multiplexing}:
In this scheme, the client or a proxy server close to the client
receives the addresses of a set
of candidate replica servers and chooses one to send the request.
Generally, this scheme imposes additional overhead in sending the set of
candidate replica servers to the client when requesting some content. Also,
due to lack of overall information, the client may choose a server with high
load, which could result in overloading servers and hence larger access latency.

\item {\em HTTP redirection}:
This is simplest and probably the least efficient means of redirecting requests.
In
this scheme, requests for content make it all the way to the origin server, at
which point the server re-directs the browser to a new URL at the HTTP protocol
level. Because the origin server or server cluster is the only point responsible
for redirecting requests, it could become a bottleneck and is quite
error prone.

\item {\em DNS indirection}:
This scheme uses Domain Name System (DNS) modifications to return the IP address
of one of a set of replica servers when the DNS server is queried by the client.
Choosing a server may depend on the location of the client. This technique is
transparent to the client.
The quality of server selection can be improved by
taking into account the server performance. Some commercial CDNs, such as
Akamai, are taking this approach.
\mycomment{uses a round robin
mechanism to allocate servers to clients because it maintains no server
performance information on which to base its selection decision.
Additionally, 
DNS redirection does not scale well because it requires the world-wide clients
of a Web site, on a DNS cache miss, to incur the long round-trip time to a
centralized name server as part of accessing the site, from wherever the client
is located.}
\mycomment{
Additionally,
DNS redirection does not scale well because when the size of the system
grows,
it may take the clients a long round-trip time to a centralized name server
to resolve a site name on a DNS cache miss. }
\mycomment{
it requires the world-wide clients
of a Web site, on a DNS cache miss, to incur the long round-trip time to a
centralized name server as part of accessing the site, from wherever the client
is located. }

\item {\em Anycasting}:
Essentially, request routing in CDN can be viewed as an application of locating
nearby copies of replicated Internet servers. Techniques such as anycasting
developed for server location can be used for request routing in CDN.
In this scheme, an \emph{anycast address/domain name}, which can be
an \emph{IP anycast address} or \emph{a URL of content}, is used to define a
group of servers that provide the same service. A client desiring to communicate
with only one of the servers sends packets with the anycast address in the
destination address field. The packet is then routed via anycast-aware
routers to at least one of the servers identified by the anycast address. This
anycast-aware routing can be integrated into the existing Internet routing
infrastructure, thereby providing request routing service to all the CDNs. In
addition, this scheme has the potential to scale up well with the growth of
the Internet.

\item{\em Peer-to-Peer Routing}:
Peer-to-peer systems are becoming widely deployed on the
Internet to disseminate content.
The participant nodes in a peer-to-peer system generally belong to different
affiliations and themselves constitute an ad-hoc network. As the network is
constantly changing, no nodes have the complete global information in time
about the network. The problem of routing requests efficiently in a distributed
manner without incurring high overhead of propagating the routing information is
a major research concern.

\end{itemize}



\section{Server Placement}

In this section, we
survey the approaches to server placement described in the literature along with
their contributions and inadequacies.
The issue here is to decide where on the Internet to place the servers,
which store replicates of objects.
The goal is to minimize the following two metrics:
the average content access latency perceived by clients and the
overall network bandwidth consumption for
transferring replicated documents from servers to clients. These two
are correlated in sense that optimizing one metric generally will
lead to the minimization of another one.

\subsection{Theoretical Approaches}
The server placement problem can be modeled as a
\emph{center placement problem}.
The center placement problem is defined as follows:
for the placement of a given number of centers, one
could consider the metric \((P_{minK})\) of minimizing the maximum
distance between a node and the nearest center.
This problem is also known as the {\em minimum K-center problem} \cite{Jamin}.
A problem
similar to the minimum K-center problem is the {\em facility location problem},
where the total cost in terms of building facilities and servicing clients
is a given constrain \cite{Qiu}.
Another theoretical approach to solve the center
placement problem based on graph theory is the
{\em k-hierarchically well-separated trees} (k-HST) \cite{Jamin, Bartal}.
We first define some notations
and then describe the k-HST and the  minimum $K$-center approaches.
\mycomment{After that,
the technique based on the facility location model is presented briefly
because of its similarity to the minimum $K$-center problem. }

\emph{Notations:} We adopt the following notations in this section: the network
is represented by a graph \(G(V,E)\), where \(V\) is the set of nodes, and
\(E \subseteq V \times V\) is the set of links. We use \(N=|V|\) to denote the 
number of nodes in $G$, and $\mathcal{T}$ to denote the number of centers we
place in the graph. We denote the distance between nodes \(u\) and \(v\) in the
graph $G$ by $d_{G}(u,v)$; we will omit $G$ when it can be deduced from the
context.

\subsubsection{\emph{k}-HST}
The {\em k}-HST algorithm consists
of two phases. In the first phase, the graph is recursively partitioned as
follows. A node is arbitrarily selected from the current (parent) partition,
which is the complete graph at the beginning,
and all the nodes that are within a random radius from this node form a new 
(child) partition. The value of the radius of the child partition is a factor
of $k$ smaller than the diameter of the parent partition. This process recurs
for each partition, until each node is in a partition of its own. It then
obtains a tree of partitions with the root node being the entire network and
leaf nodes being individual nodes in the network. In the second phase, a virtual
node is assigned to each of the partitions at each level. Each virtual node in
a parent partition becomes the parent of the virtual nodes of the child
partitions. The length of the links from a virtual node to its children is
half of the partition diameter.
Together, the virtual nodes also form a tree.
Figure \ref{hst:fig} gives an example of generating a 1-HST tree based on
the algorithm described above from a graph. All the links on the graph
have the length of one.

\myfig{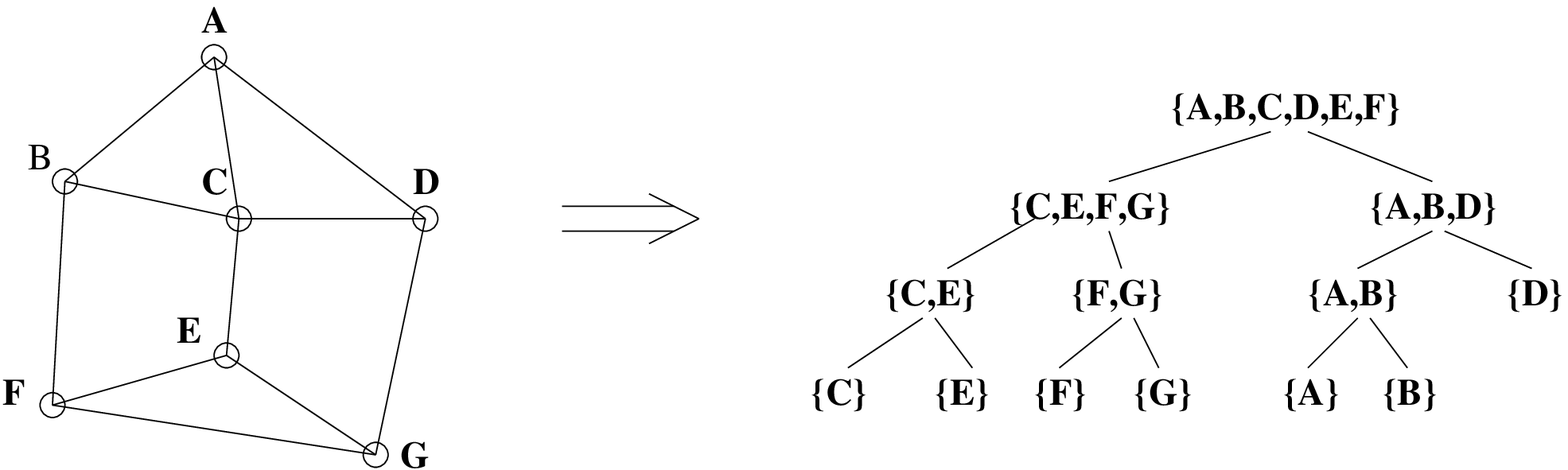}{4.2 in}{Example of Generating {\em 1}-HST}

The randomization of a partition radius is done so that the probability of a 
short link being cut by partitioning decreases exponentially as one climbs the
tree. Hence nodes close together are more likely to be partitioned lower down
the tree. Taking advantage of this characteristics of the resulting $k$-HST
tree, the following greedy algorithm can be devised to find the number of
centers needed when the maximum center-node distance is bounded
by $\mathcal{D}$.

Let node $r$ be the root of the partition tree, $\mathcal{N}_{i}$ be the
children of node $i$ on the partition tree, and $\mathcal{L}$ be a list of
partitions sorted in the decreasing order of the partition diameter at all
times. $H_{\mathcal{L}}$ denotes the partition at the head of the list, and
$diam(H_{\mathcal{L}})$ its diameter. The following shows the greedy algorithm
on the $k$-HST tree.
{\tt
\\ $\mathcal{L} \leftarrow \mathcal{N}_{r}$ \\
while( $diam(H_{\mathcal{L}}) > \mathcal{D})$ \\
begin \\
\indent
$h \leftarrow H_{\mathcal{L}}$ \\
\indent
$\mathcal{L} \leftarrow \mathcal{L} - H_{\mathcal{L}}$ \\
\indent
$\mathcal{L} \leftarrow \mathcal{L} \cup \mathcal{N}_{h}$ \\
end \\
}
The algorithm pushes the centers down the tree until it discovers a partition
with diameter $\leq \mathcal{D}$. The number of partitions, $|\mathcal{L}|$, 
is the minimum number of centers required to satisfy the performance metric
$P_{diam}$. To select the actual centers, it simply needs to
set the virtual nodes of these partitions in $\mathcal{L}$ to be the centers.
Using the example of Figure \ref{hst:fig}, if we set $\mathcal{D}$ as $2$,
we will get the partition of $\{\{C,E,F,G\}, \{A,B,D\}\}$ after running the
above greedy algorithm. This result shows that if we need to ensure that the
distance between any single node and the closest server to be no greater than
two,
we have to deploy two servers with each located within node set $\{C,E,F,G\}$
and $\{A,B,D\}$, respectively.

The $k$-HST-based greedy placement algorithm presented above can not only
tell us the
number of centers needed to satisfy the performance metric $P_{diam}$,
but it can also be used to determine
their placement for any given budget of centers.
For example, to place $K$ centers, it simply needs to change
line 2 in the above algorithm with "while( $|\mathcal{L}| < K$ )".
Obviously, the performance metric $P_{diam}$ may no longer be satisfied
for $K$ below a certain number.

\subsubsection{Minimum \emph{K}-center}
The minimum $K$-center problem is NP-complete \cite{Garey}.
However, if we are willing
to tolerate inaccuracies within a factor of 2, i.e.,
the maximum distance between
a node and the nearest center being no worse than twice the maximum in the
optimal case, the problem is solvable in $O(N|E|)$ \cite{Vazirani} as follows.

Given a graph $G=(V,E)$ and all its edges arranged in non-decreasing order by
edge cost, $c: c(e_{1}) \leq c(e_{2}) \leq ... \leq c(e_{m})$, let
$G_{i} = (V,E_{i})$, where $E_{i} = \{ e_{1},e_{2},...e_{i}\}$. A {\em square
graph} of $G$, $G^{2}$ is the graph containing $V$ and edge $(u,v)$ wherever
there is a path between $u$ and $v$ in $G$ of at most two hops, $u \neq v$ ---
hence some edges in $G^{2}$ are pseudo edges, in that they do not exist in $G$.
An {\em independent set} of a graph $G=(V,E)$ is a subset $V' \subseteq V$ such
that, for all $u,v \in V'$, the edge $(u,v)$ is {\em not} in $E$. An independent
set of $G^{2}$ is thus a set of nodes in $G$ that are at least three hops apart
in $G$. We also define a {\em maximal} independent set $M$ as an independent
set $V'$ such that all nodes in $V - V'$ are at most one hop away from nodes
in $V'$.

The outline of the minimum $K$-center algorithm from \cite{Vazirani} is shown as
follows:
\begin{enumerate}
\item Construct $G_{1}^{2}$, $G_{2}^{2}$, ... ,$G_{m}^{2}$
\item Compute $M_{i}$ for each $G_{i}^{2}$
\item Find smallest $i$ such that $|M_{i}| \leq K$, say $j$
\item $M_{j}$ is the set of $K$ center.
\end{enumerate}

The basic observation is that the cost of the optimal solution to the $K$-center
problem is the cost of $e_{i}$, where $i$ is the smallest index such that
$G_{i}$ has a dominating set\footnote{A dominating set is a set of
$\mathcal{D}$ nodes such that every $v \in V$ is either in $\mathcal{D}$ or
has a neighbor in $\mathcal{D}$.} of size at most $K$. This is true since the
set of center nodes is a dominating set, and if $G_{i}$ has a dominating set
of size $K$, then choosing this set to be the centers guarantees that the
distance from a node to the nearest center is bounded by $e_{i}$. The second
observation is that a star topology in $G_{i}$ transfers into a clique 
(full-mesh) in $G_{i}^{2}$. Thus, a maximal independent set of size $K$ in
$G_{i}^{2}$ implies that there exists a set of $K$ stars in $G$, such that the
cost of each edge in it is bounded by $2e_{i}$: the smaller the $i$, the
larger the $K$. The solution to the minimum $K$-center problem is the
$G_{i}^{2}$ with $K$ stars. Note that this approximation does not always
yield a unique solution.

The 2-approximation minimum $K$-center algorithm can also be used to determine
the number of centers needed to satisfy the performance metric $P_{diam}$ by
picking an index $k$ such that $c(e_{k}) \leq \mathcal{D} / 2$. The maximum
distance between a node and the nearest center in $G_{k}$ is then at most
$\mathcal{D}$, and the number of centers needed is $|M_{k}|$.

\mycomment{
\subsubsection{Facility Location Problem}
Given a set of locations
at which facilities (centers) may be built, building a facility at location
\(i\) incurs a cost of \(f_{i}\). Each client \(j\) must be assigned to one
facility, incurring a cost of \(d_{j}d_{G}(i,j)\), where \(d_{j}\) denotes the
demand of the client \(j\). \mycomment{
and \(c_{ij}\) denotes the distance between \(i\) and
\(j\). }
The objective is to find a solution (i.e., both the number of facilities
and the locations of the facilities) of the minimum total cost.

There have been a number of approximation algorithms developed for this NP-hard
problem in the metric space. A \emph{$\rho$-approximation
algorithm} is a polynomial-time algorithm that always finds a feasible solution
with an objective function value within a factor $\rho$ of the optimal solution.
The best approximation algorithm known today was developed by Charikar and Guha 
\cite{Charikar}, who gave a 1.728-approximation algorithm. }


\subsection{Heuristic Solutions}

The theoretical solutions described above
are either computationally expensive or do not consider the characteristics
of the network and workload. Thereby, they are very difficult to apply in
practice \cite{Radoslavov} and may not be suitable for the real CDN systems.
Some heuristic or suboptimal algorithms were
proposed, which leverage some existing information from CDNs,
such as the work load pattern and the network topology 
\cite{Krishnan,Qiu,Jamin1,Radoslavov}, and offer adequate 
solution with much lower computation complexity.

\subsubsection{Greedy Algorithm}
A greedy algorithm was proposed by P. Krishnan et al. \cite{Krishnan} for the
cache location problem.
Qiu et al. \cite{Qiu} adapted this algorithm for the server placement problem in
CDN.

The basic idea of the greedy algorithm is as follows. Suppose it needs to choose
$M$ servers among $N$ potential sites. It chooses one site at a time. In
the first iteration, it evaluates each of the $N$ potential sites individually
to determine its suitability for hosting a server. It computes the cost
associated with each site under the assumption that accesses from all
clients converge at that site, and picks the site that yields the lowest cost,
e.g., the bandwidth consumption.
In the second iteration, it searches for a second site that, in
conjunction with the site already picked, yields the lowest cost. In general,
in computing the cost, the algorithm
assumes that clients direct their accesses to the 
nearest server, i.e., one that can be reached with the lowest cost.
The iteration continues until $M$ servers have been chosen.

Jamin et al. \cite{Jamin1} discussed a more general algorithm:
$\ell$-backtracking greedy algorithm. It differs from the basic one as
follows: in each iteration, the $\ell$-backtracking greedy algorithm
checks all the possible combinations achievable
by removing $l$ of the already placed servers and replacing them with
$\ell + 1$ new servers. Thus, the basic greedy algorithm is 
0-backtracking greedy algorithm.

In evaluating the relative performance of the greedy replica placement
strategy to that of the optimal ones, Qiu et al. \cite{Qiu}
extracted the client locations from the workload
traces from MSNBC, ClarkNet and NASA Kennedy
and reduced the number of the locations to the scale of thousands 
by clustering the clients that
are topologically close together. They utilized two types of network topologies:
one by simulating the network as random graph and another one by
deriving the real topology
from BGP routing table. Under this experimental configuration, 
the greedy algorithm
performs remarkably well (within a factor of 1.1-1.5) 
compared to the computationally expensive optimal solution and the computation
needed is several magnitudes less.
Also, the greedy algorithm is relatively insensitive to imperfect input data.
Unfortunately, this greedy placement requires knowledge of the client
locations in the network and all pairwise inter-node distances. This
information in many cases may not be available.

\subsubsection{Topology-informed Placement Strategy}
A topology-informed placement strategy, called "Transit Node",
was first discussed by Jamin et al. \cite{Jamin1}. It works as follows.
Assuming that nodes with the highest outdegrees\footnote{The outdegree of a node
is the number of other nodes it is connected to.} can reach more nodes with
smaller latency, we place servers on candidate hosts in descending
order of outdegrees. This is called \emph{Transit Node} heuristic under the
assumption that nodes in the core of the Internet transit points will
have the highest outdegrees. \mycomment{Their result shows that the
transit node heuristic can perform almost as well as the greedy placement.
However,} Due to the lack of more detailed network topology, it uses only
Autonomous Systems (AS) topologies where each node represents a single AS, and
node link corresponds to AS-level BGP peering.
An improved topology-informed placement strategy is proposed by Radoslavov et
al \cite{Radoslavov}. They leverage the router-level Internet topology, instead
of only AS-level topology. In this strategy, each LAN associated with a router
is a potential site to place a server, rather than each AS being a site.
\mycomment{
Their finding shows that using router-level topology information results in
better performance than that achieved by only exploiting AS-level topology
knowledge.}

\mycomment{
Compared to the greedy algorithm, the topology-informed placement strategy
does not need the detailed knowledge of the network topology and
the expected client locations on the topology, and can achieve similar
performance result as the greedy algorithm does.
}

In the experiments of both strategies,
the network topology is extracted from the Internet and
the set of client locations is derived from some Web site trace log.
The result shows that the
transit node heuristic can perform almost as well as the greedy placement,
and that using router-level topology information results in
better performance than that achieved by only exploiting AS-level topology
knowledge. Also, they found that the performance improvement diminishes
when increasing the number of the servers and only explored
the performance of up to tens of servers.

In practice, due to the capacity
limit of a single site, a CDN may consist of much more servers, e.g.,
Akamai deploys around 13,000 servers in 64 countries \cite{Akamai}. To
the best of our knowledge, the problem of optimizing the placement of
such a large number 
of servers on the Internet is not well explored.

\section{Request Service}
Two steps of operations are involved in servicing a request in a CDN:
locating a suitable server holding the replicates of the requested object,
which we name as {\em server location},
and redirecting the request to the selected server,
which is called {\em request routing}.
In a totally distributed request routing system, such as 
anycasting, the request is forwarded directly to the proper
servers via anycast-aware routers, in which these two
operations are combined together indeed. In this section, we first investigate
these two issues and then explore the known anycasting schemes in depth.
As one special type of CDN with no infrastructure support, peer-to-peer systems
have been shed spotlight, either from academic or industry.
We will also discuss the request routing problem in these systems in the last
subsection.

\subsection{Server Location}

\myfig{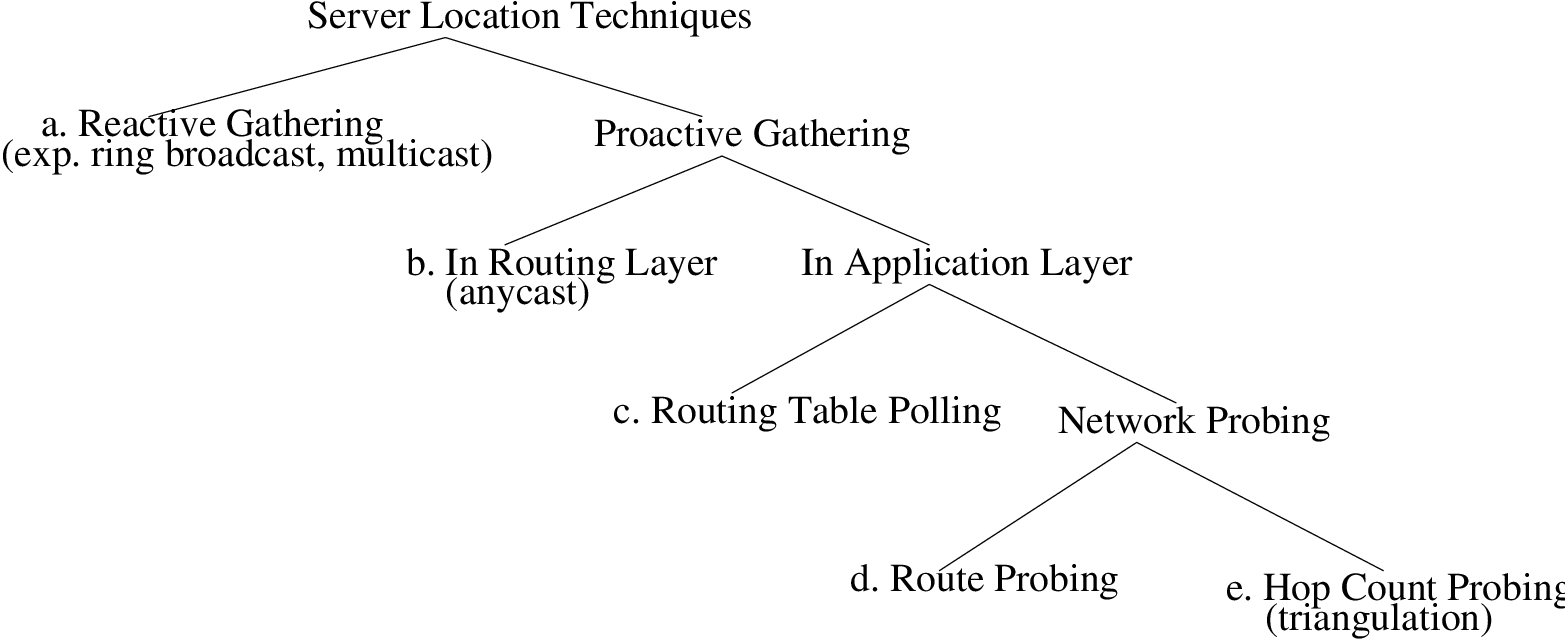}{4 in}{Server Location Techniques}
There are a number of possible ways to locate a nearby or suitable server
in an internetwork environment, as summarized in
Figure \ref{server_loc:fig} \cite{Guyton}. The first choice about server
location strategy is whether server
location information is gathered in {\em reaction} to client requests for
nearby servers, e.g., using a multicast scheme, or whether this
information is gathered {\em proactively}. The next choice is whether support
should be provided by the routing layer. Such support can reduce gathering
costs, but given the practical difficulties of implementing widespread
router changes,
we also consider techniques to gather server location information in the
application layer. Finally, we compare the cost of polling routing table against
gathering information via network probes.

\paragraph{Reactive vs. proactive gathering}
Broadcast and multicast are two typical reactive approaches to locate
suitable servers. Boggs \cite{Boggs} proposed an
{\em expanding ring} broadcast mechanism
that iteratively enlarges concentric rings around
the broadcasting host. More recent approaches have considered
various forms of multicast routing. That is, given a multicast group joined
by all instances of a particular type of server, one can choose the server
that responds most quickly to a group multicast message. Since each time
a client wants to locate a server providing a particular service, a message
has to be multicasted to the entire group of servers, this approach 
wastes the precious network bandwidth. 
In contrast, in a proactive scheme, some agents, such as routers
or dedicated applications, gather the network and server information by
sending probe messages to the candidate servers or collect load information
dispatched by servers, and maintain the server location or load database.
Clients that
request some service can locate a proper server providing the desired service
by only sending a query to the agents. This way, the unnecessary message 
transmission can be avoided.

\paragraph{Routing layer vs. application layer}
Partridge et al. \cite{Partridge} proposed an {\em anycasting} mechanism,
particularly in IP layer, which attempts to deliver a request to
one nearby server.
Anycasting is appealing because it can avoid burdening links with repeated
requests to gather server distance information.
However, a downside is that IP
anycasting assumes that all servers provide equal service\footnote{
Partridge et al. \cite{Partridge}
specifies that a single server is selected in response to an anycast request,
and that any of the servers in the anycast group are equally usable.}. It
therefore cannot handle server quality differences without programming
policy constraints into routers. In contrast to IP anycasting,
an application-level
location service could include server quality into the selection criterion, 
after a handful of nearby servers have been selected from the
database. As an additional practical advantage, an application-level service
could be implemented without routing support, albeit at higher network cost.
The main disadvantage of building the server location database at the
application-level is that doing so requires periodic updating. At the routing
layer the location database can be constructed incrementally by monitoring
routing update traffic and group join/leave requests. Some approaches were
proposed which aim to enjoy the benefits of both. Wood et al. \cite{Wood}
built an
application-level location service that constructs the server location
database by monitoring the routing-layer traffic. Fei et al. \cite{Fei}
proposed an application-level anycasting mechanism to provide server
location service. In their system, only when the load on a server changes
{\em significantly}, the server pushes the changed status to the agents and
then triggers updating the database so as to cut down the updating operation
cost.

\paragraph{Polling routing table vs. network probing}
By polling routing tables, we can build a
connectivity graph from a measurement beacon to one server by retrieving the
local routing table, determining the next hop along the path, retrieving the
routing table for that router, and so on, until we reach the destination.
We can extend this algorithm to discover routes to all servers by iterating
over the servers, taking care not to retrieve a routing table that has
already been retrieved.
In network probing approach,
some measurement servers are responsible to explore the route to each of the
replica servers by probing the servers. When a client asks one of the
measurement servers for a list of nearby replica servers, a measurement
server explores the route back to the client and adds that information to
its connectivity database. It then searches the databases for servers near
the client.



\subsection{Request Routing}
The request routing schemes proposed so far fall into the five categories:
{\em client multiplexing}, {\em DNS indirection}, {\em HTTP redirection},
{\em anycasting} and {\em peer-to-peer routing}.
We will discuss the former three in the remaining part of this subsection and
the latter two in the other two subsections.

\subsubsection{Client Multiplexing}
In this approach, the client (Web browser or a proxy server) obtains
the addresses of a set of
physical replica servers and chooses one to send its request to. Three main
mechanisms belong in this category.

The first one is that the DNS server of the service provider returns the IP
addresses of servers holding a replica of the object. The client's DNS resolver
chooses a server among these. To decide, the resolver may issue probes to the
servers and choose based on response times to these probes, or it may collect
reports from the clients on performance of past accesses to these servers.
This approach is used by Bhattacharjee et al. \cite{Bhattacharjee}
and by Beck and Moore \cite{Beck}. Its advantage is that no extra communication
is added to the critical path of request processing. There are also several
shortcomings. First, the approach relies on clients using a customized DNS
resolver. If this resolver relies on client performance reports, then the
client (i.e., browser or proxy) software must be modified as well. Second,
the DNS infrastructure relies heavily on DNS response caching to cope with
its load. Therefore, replica server sets cannot be changed frequently
without the
danger of resolvers using stale replica server sets. At the same time, reducing
the caching time may incur DNS queries more frequently and hence just moves the
stability bottleneck to the DNS
infrastructure, provided enough clients adopt the approach.\mycomment{
Thus, this approach
is most appropriate for static replication. }

The second approach relies on Java applets to perform replica server selection
on the client \cite{Yoshikawa}. The URL of an object actually points to a Java
applet, which embeds the knowledge about the current replica server set and the
procedure for replica server selection.
This approach requires no changes to clients.
However, unlike the previous approach, it involves an extra TCP communication
to download the applet.

The third approach, proposed by \cite{Baentsch},
 propagates information about replica server
sets in HTTP headers. It requires changes to both Web servers and clients 
(proxy servers in this case) to process extra headers. Clients must also be
modified to implement replica server selection.

\mycomment{\subsubsection{IP Multiplexing}
A plethora of commercial products offers multiplexing routers. In this approach,
a special multiplexing router (or a multiplexer) is placed in front of a
server farm. The domain name of the Web site is mapped to the IP address of
this router, so all clients send their packets to it. When the first packet
from a client arrives, the multiplexer selects a server in the farm and
forwards the packet (and all subsequent packets in the same session) to this
server. The multiplexer also intercepts the response packets and modifies
their headers to contain its IP address rather than that of the server. The
multiplexer maintains a {\em sessions database}, which records a server chosen
for each active communication session. The sessions database is needed to ensure
that, once a server is chosen for servicing a request, all packets from this
client in this communication session go the same server.

This approach uses standard clients and DNS and Web servers. However, by
having a focal communication point, it does not scale geographically. Thus,
it is mostly used for load sharing in a server farm on a local-area network. }

\subsubsection{DNS Indirection}
Several domain name server implementations allow the Web site's DNS server
to map a host domain name to a set of IP addresses and choose one of them for
every client query, based on such factors as the query origin and the load
of replica servers. The difference with DNS-based client multiplexing is
that choosing a replica server occurs at the Web site or the DNS infrastructure,
not at the client's DNS
resolver. \mycomment{This approach can scale well geographically.}
Unfortunately,
DNS response caching by clients complicates changing replica server sets and
controlling request distribution among replica servers.
At the same time, reducing
the lifetime of cached DNS responses may shift the performance bottleneck
to the DNS infrastructure.
In general, DNS system was designed for mostly an append-only database of
existing mappings between a host name and an IP address that
rarely ever changes.
\mycomment{
Dynamic replication schemes, however, constantly change the
set of replicas for a given object. }


\subsubsection{HTTP Redirection}

HTTP protocols allow a Web server to respond to a client request with a
special message that tells the client to re-submit its request to another
server. This mechanism can be used to build a special Web server which accepts
client requests, chooses replica servers for them and redirects clients to these
servers. Commercially, Cisco Distributed Director \cite{Cisco} and
WindDance Web Challenger \cite{WindDance} implemented this functionality.


An advantage of HTTP-level redirection is that replication can be managed
at fine granularity, down to individual Web pages, whereas other mechanisms
postulate the entire Web site as the granule. A disadvantage is that this
mechanism is quite heavyweight. Not only does it introduce an extra message
round-trip into request processing, but also this round-trip is done over HTTP,
which uses the expensive TCP protocol as the transport layer.


\mycomment{
\subsection{Anycasting}

As defined \cite{Partridge}, anycasting provides:
``a stateless best effort delivery of an anycast packet to at least one host,
and preferably only one host, which serves the anycast address". The anycasting
schemes proposed in literature so far can be classified into two classes:
IP anycasting and application-layer anycasting.
}

\subsection{Anycasting}

As defined \cite{Partridge}, anycasting provides:
``a stateless best effort delivery of an anycast packet to at least one host,
and preferably only one host, which serves the anycast address". The anycasting
schemes proposed in literature so far can be classified into two classes:
IP anycasting and application-level anycasting.

\subsubsection{IP Anycasting}

\mycomment{In the next generation of the Internet routing protocol (IPv6),
an {\em anycast}
service \cite{} will be supported. This service assume that the same IP address
is assigned to a set of hosts, and each IP router has in its routing table
a path to the host that is the closest to this router. Thus, different IP
routers have paths to different hosts with the same IP address.

There is virtually no overhead for request indirection with anycast, since
packets travel along router paths anyway. However, it assumes very static
replica sets (since a change in the replica set would take long time to 
reflect in router tables throughout the Internet) and a rigid request
distribution (because all requests from a given client normally go to the same
host, chosen based on the network proximity). Advances in research on
{\em active networks} \cite{}, where applications can inject ad-hoc programs
into network router, may alleviate the last limitation. }

\myfig{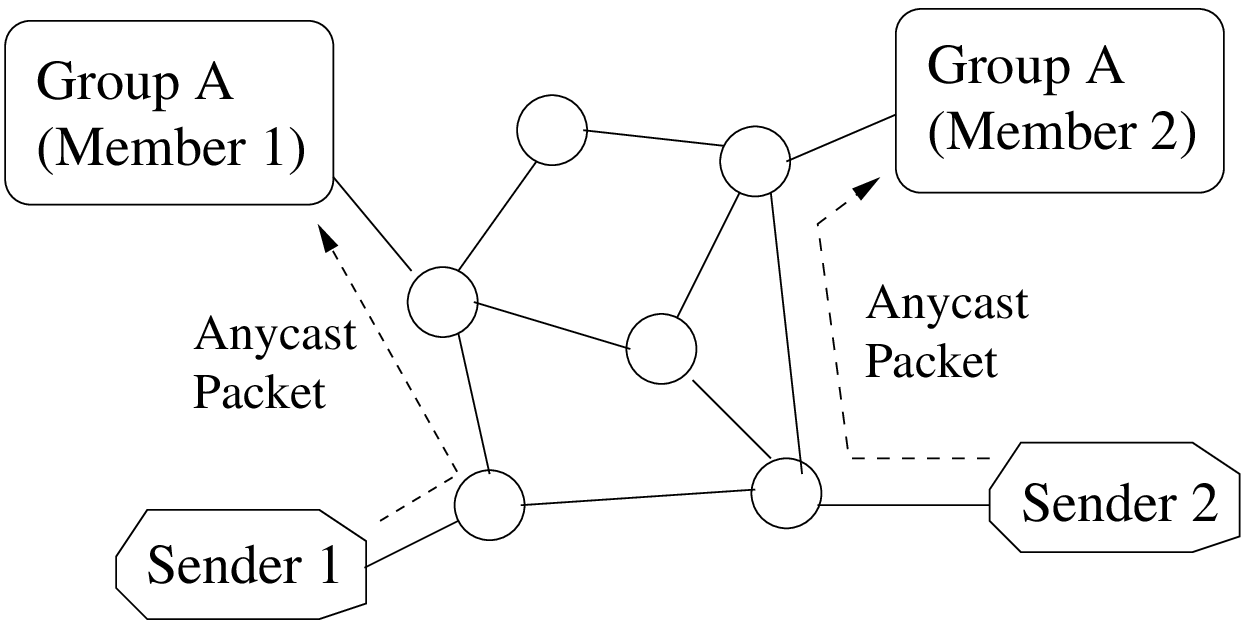}{3 in}{Illustration of IP Anycast}

IP anycasting was proposed by Partridge et al. \cite{Partridge}. Such service
assumes that the same IP address is assigned to a set of hosts, and each
IP router has in its routing table a path to the host that is the
closest\footnote{'closest' is defined according to the routing system's
measure of distance.} to
this router. Thus, different IP routers have paths to different hosts with
the same IP address. Figure \ref{IP_anycast:fig} illustrates IP anycasting.

The traditional approach routes IP anycast addresses using the unicast routing
protocols, a design decision that makes IP anycast unscalable.
The anycast group topology may not be hierarchical or comply with the unicast
topology and thereby routing anycast packets using the unicast routing
protocols requires advertising each global anycast address separately. This
requirement causes the routing tables to grow proportionally to the
number of all global anycast groups in the entire Internet, and hence does
not scale.

\myfig{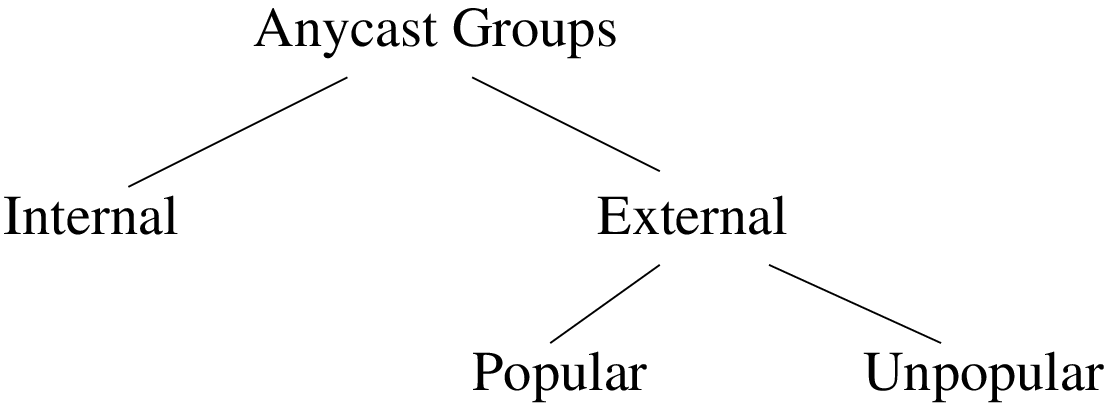}{3 in}{Anycast Group Classification at An Edge Domain}

Katabi et al. \cite{Katabi} proposed a framework for scalable global IP
anycast (GIA).
The framework
scales by capturing the special characteristics of the anycast service in its
inter-domain routing protocol, which generates two types of routes:
default inexpensive routes that consume no bandwidth or storage space,
and enhanced shortest path routes that are customized according to the
beneficiary domain's interests.
Figure \ref{IP_anycast_class:fig} shows the anycast group classification at an
edge domain.

For routing internal anycast groups, GIA uses
the unicast intra-domain routing protocol,
either based on the distance-vector algorithm, e.g., RIP, or on the link-state
algorithm, such as OSPF. This approach stays scalable because the number of
internal group is controllable by the domain itself. To route an
unpopular anycast group, GIA uses a default route. A default route is determined
by the unicast prefix of the home domain,
which is part of the anycast address, and does not consume any bandwidth
to be generated and does not need any storage space in the routing
tables.
A router that receives an anycast packet addressed to an unpopular anycast
group forwards the packet to the group member in the home domain.
For popular anycast groups, GIA generates shortest path routes and forwards
the packet address to a popular anycast group via the shortest path route.

Although IP anycast is suitable for request routing and service location,
it has the following limitations:
\begin{itemize}
\item Some part of the IP address space must be allocated to anycast address.
\item Anycast addresses requires router support.
\item The selection of the server to which an anycast packet is sent is
made entirely within the network with no option for user selection or input.
\item Consistent with the stateless nature of IP, the destination is
determined on a per-packet basis.
\end{itemize}

\subsubsection{Application-Level Anycasting}

Since the network layer is able to effectively determine the shortest path,
it is well suited for the IP anycasting service that selects the closest
server based upon a shortest path metric such as hop count. On the other
hand, an application layer approach is better suited at handling a
variety of other metrics such as server throughput.
Here we survey some typical systems in this field, which we call content
routing network.


Fei et al. \cite{Fei} observed the shortcomings of IP anycast and proposed
a framework for application-level anycasting service. In their design, the
service consists of a set of {\em anycast resolvers}, which performs the
{\em anycast domain names} (ADN) to IP address mapping. Clients interact with
the anycast resolvers by generating an anycast query. The resolver processes
the query and replies with an anycast response. A key feature of the system
is the presence of a metric database, associated with each anycast resolver,
containing performance data about replica servers.
The performance data can be used
in the selection of a server from a group, based on user-specified performance
criteria. It estimates the server performance 
with manageable overhead by combining server pushes with client probes. However,
deploying such a system requires the changes to the servers as well as
the clients, which is prohibitively costly considering the possibly huge 
number of servers and clients.

\myfig{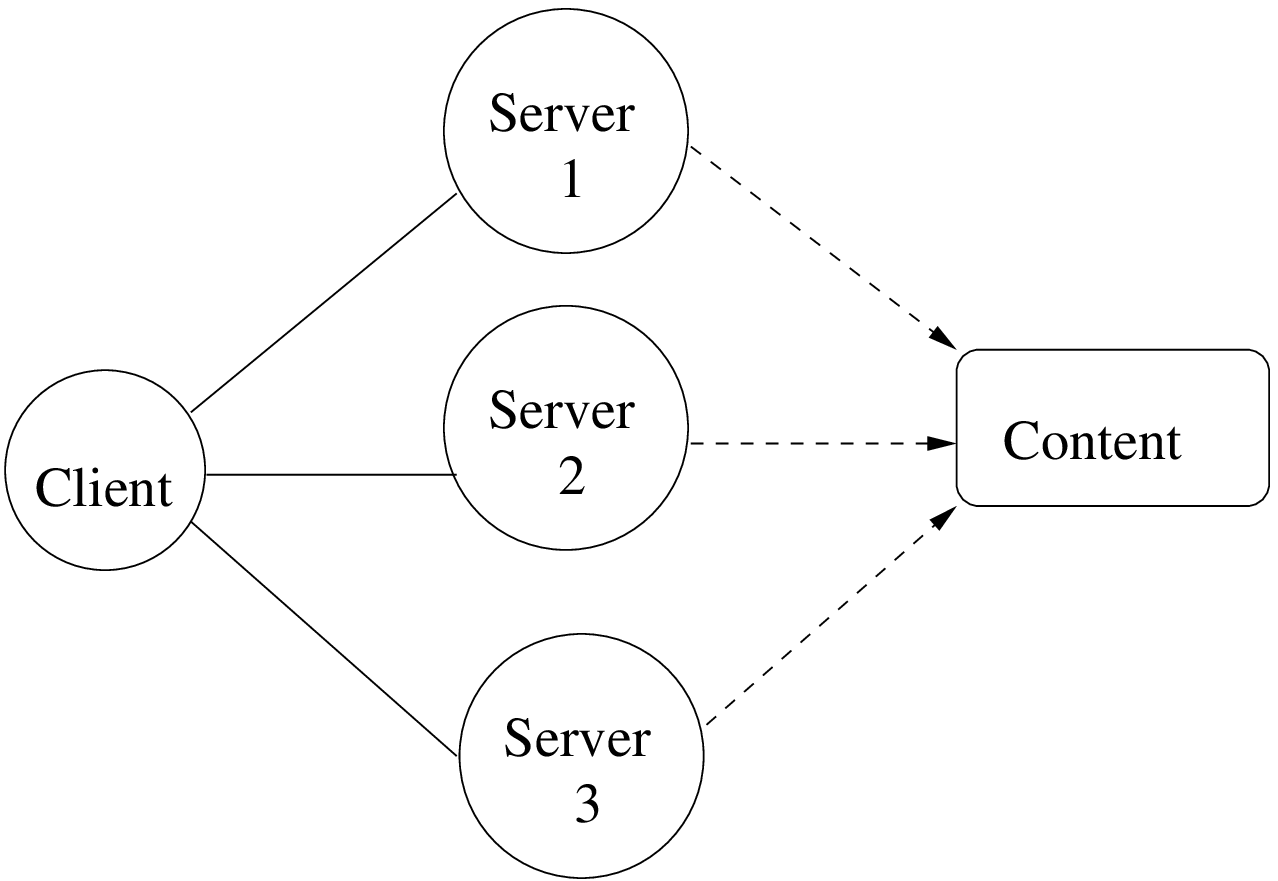}{2.2 in}{Content Layer Routing}

Adjie-Winoto et al. \cite{Adjie-Winoto} proposed an intentional naming
system (INS), which is designed as a resource discovery and service
location system for
dynamic and mobile networks of devices and computers. INS applications may be
{\em services} or {\em clients}: services provide functionality or data and
clients request and access these. {\em Intentional Name Resolvers (INRs)} form
an application-level overlay network to exchange services descriptions,
construct a local cache based on these advertisements and route
client requests to the appropriate services. INS uses a simple language based
on attributes and values for its names, which enable the service to be
specified precisely. In addition, INS implements a late binding
mechanism that integrates name resolution and message routing, enabling
clients to continue communicating with end-nodes even if the name-to-address
mappings change while a session is in progress. However, INS is not designed
to provide global reachability information, and the attribute-based naming 
is less scalable than a hierarchical namespace provided by URL. Also, INS's
late binding, where every message packet contains a name, is too expensive to
use for content distribution.


Gritter et al. \cite{Gritter} designed a framework for content routing
support in the Internet. Replica servers can be viewed as offering
alternate routes to access the content that the client requests, as depicted in
Figure \ref{content_routing:fig}. Network-integrated content routing provides
support in the core of the Internet to distribute, maintain and make use of
information about content reachability. This is performed by routers that
are extended to support naming. They designed the {\em Name-Based Routing
Protocol (NBRP)}, which performs routing by name with a structure similar to
BGP. Like BGP, NBRP is a distance-vector routing algorithm with path
information: an NBRP routing advertisement contains the path of content routers
toward a content server. An {\em Internet Name Resolution Protocol (INRP)} is
developed to perform efficient lookup on the distributed integrated
name-based routing system.


Clients that desire some content initiate content request by contacting a local
content router. Each content router maintains a set of name-to-next-hop
mappings, just as an IP router maps address prefixes to next hops. When an
INRP request arrives, the desired name is looked up in the name routing table,
and the next hop is chosen based on the information associated with the known
routes. The content router forwards the request to the next content router, and
in this way the request proceeds toward the ``best" content server, as shown in
Figure \ref{inrp:fig} (The routing information kept for a name is typically
just the path of content routers to the content server, although it may be
augmented with load information or metrics directly measured by a content
router). When an INRP request reaches the content router adjacent
to the ``best" content server, that router sends back a response message
containing the address of the preferred server. This response is sent back
along the same path of content routers. If no response appears, intermediate
content routers can select alternate routes and retry the name lookup.
In this fashion, client requests are routed over the best path to the desired
content. INRP thus provides an ``anycast" capability at the content level.

\myfig{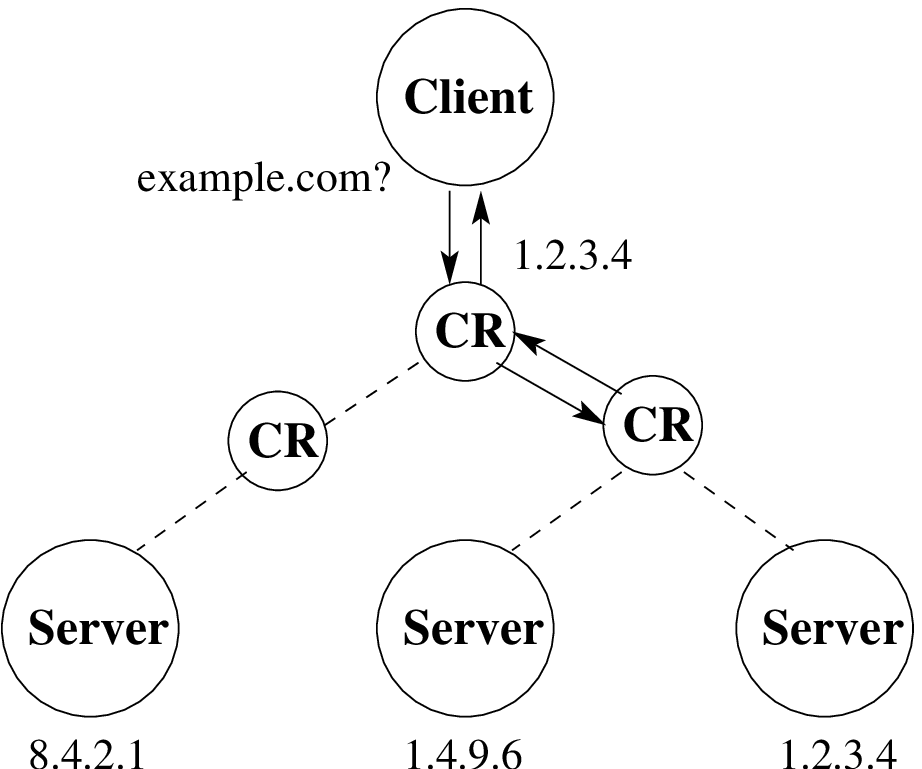}{2.0 in}{Internet Name Resolution Protocol}

\subsection{Peer-to-Peer Systems}
Peer-to-peer systems build the information retrieval network on the members
themselves instead of relying on a dedicated infrastructure like the
traditional CDNs do. As a result, peer-to-peer systems are more fault-tolerant
than the common CDNs.
Also, peer-to-peer
systems are more suitable for content producers who are individuals, who may
not be able to access or afford the common CDNs, most of which are commercial
ones.

Ian Clarke et al. \cite{clarke00freenet}
designed Freenet, a distributed adaptive peer-to-peer system that
enables the storage and retrieval of data while maintaining the anonymity
of readers and authors. Freenet operates as a peer-to-peer system where nodes
request a file store or retrieve service to their immediate neighbors using
a location-independent naming key. Requests are forwarded hop-by-hop,
a way similar
to IP routing, and have a limited hops-to-live as well as a unique random id
number to avoid loops in routing. Each node has a data store to which it must
allow network access, as well as a dynamic routing table with keys associated
with node addresses. File keys are generated using hash functions. Each file
has a random public/private key pair to serve as a namespace called
signed-subspace key (SSK) and a keyword-signed key (KSK) generated by a short
descriptive text. A user publishes his descriptive string and subspace
public key, and keeps his private key secret so that
no other ones can add files to his
subspace. A content-hash key is useful for updating and splitting files
since the old version of the file remains temporarily available while a new
version is being added to the system. There is more than one solutions proposed
to search for keys, including insertion of indirect files by users with
pointer to the real files, and publicizing public key compilations by users.
Once the file key is known, a user will ask its node to retrieve the file. This
node will check its own data store, and use its routing table to forward the
request to a neighboring node if it does not have a copy of the file. Requests
propagate as a steepest-ascent hill-climbing search with backtracking until
the file is found or the request times out. A similar search is performed to
insert new files. Similar keys are located and success is returned if the
hops-to-live is reached without any collisions. Essentially, a trend will
form where nodes become experts on similar keys located on the
same node, and caching brings copies of file closer to the requester. When
the system starts running out storage, the least recently used files are
replaced. The system achieves security by enforcing anonymous requesters
and senders, as well as employing a cryptographic protocol to add new nodes to
the system. Freenet does not assign responsibility for documents
to specific servers; instead, its lookups take the form of searches for
cached copies. This allows Freenet to provide a degree of anonymity, but
prevents it from guaranteeing retrieval of existing documents or from
providing low bounds on retrieval costs.

\myfig{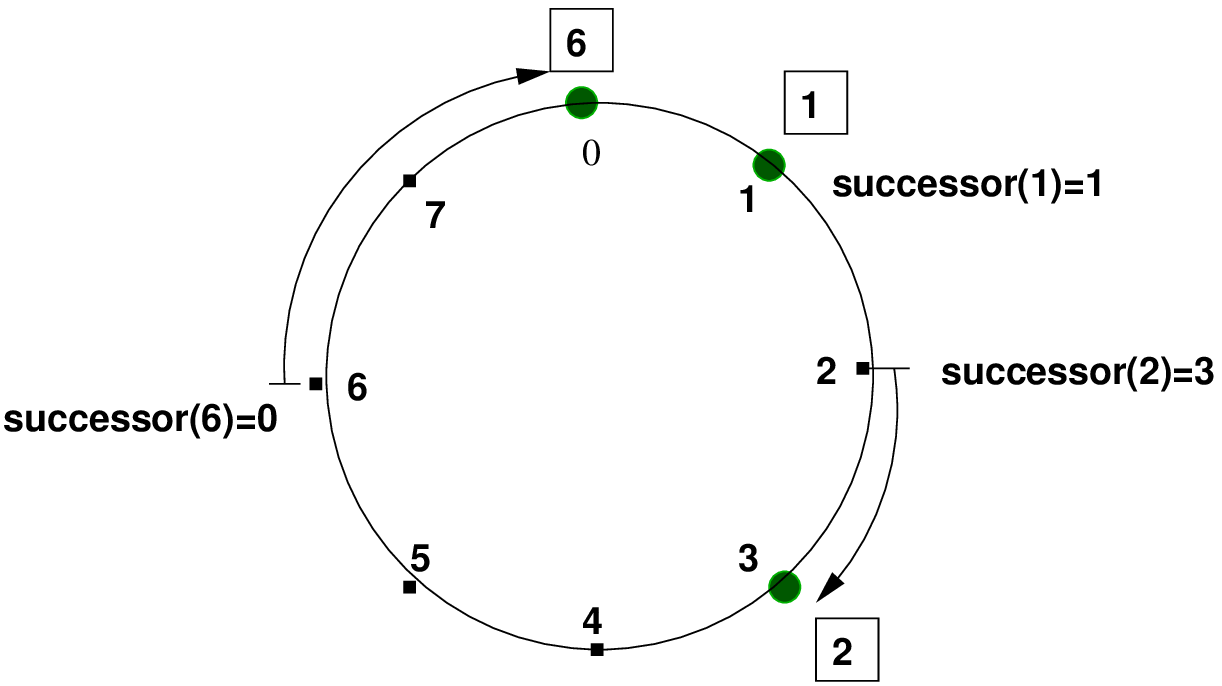}{3 in}{An identifier circle consisting of the three nodes
0, 1, and 3. In this example, key 1 is located at node 1, key 2 at node 3,
and key 6 at node 0.}

Ion Stoica et al. \cite{Stoica} presented the Chord protocol that is designed to
map a key onto a node in a distributed, peer-to-peer network
whose size and composition change intermittently. Chord is proposed to meet
the challenges faced by large-scale peer-to-peer network, namely, load balance,
decentralization, scalability, availability and flexible naming. The Chord
protocol employs the consistent hashing function to acquire two $m$-bit
identifiers for the node and the key, respectively. The identifiers are ordered
in an identifier circle modulo $2^{m}$ ($m$ is in the range of $O(logN)$ and
$N$ is
the number of nodes in the network), using which the keys may be assigned
to the nodes with the successor node technique, as shown in Figure
\ref{chord_hash:fig}. Consistent hashing is designed
to let nodes enter and leave the network with minimal disruption. Further more,
it is proven that each node is responsible for at most $(1+\epsilon)K/N$ keys,
where $K$ is the total number of keys.
Each node, $n$, maintains a routing table with (at most) $m$ entries, called
the {\em finger table}. The $i^{th}$ entry in the table at node $n$ contains
the identity of the {\em first} node, $s$, that succeeds $n$ by at least
$2^{i-1}$ on the identifier circle, i.e., $s=successor(n+2^{i-1})$, where
$1 \leq i \leq m$. When a node $n$ looks for the node holding key $k$,
it searches
its finger table for the node $j$ whose ID most immediately precedes $k$, and
asks $j$ for the node it knows whose ID is closest to $k$. By repeating this
process, $n$ learns about nodes with IDs closer and closer to $k$. Eventually,
node $n$ finds the node holding $k$ in $O(logN)$ steps.
Also, Chord is designed to ease the additions and withdrawals
of nodes, and the main idea for achieving this is to find the new predecessor
and successor nodes on the identifier circle after addition/withdrawal
of nodes in the network. When multiple nodes join or fail, a stabilizing routine
is executed in the Chord in order to remap keys to the nodes so that the
integrity of the identifier circle is maintained. In a word, Chord features
simplicity, provable correctness and provable performance even in the face of
concurrent node arrivals and departures. However, Chord needs specific
mechanism to heal partitioned rings and security mechanism to tackle
safety issues, e.g., malicious or buggy set of participants,
which may present an incorrect view of
the Chord ring. Also, anonymity of participant is not taken into account
in Chord.

\section{An Example of CDN -- Akamai}
Akamai \cite{Akamai} is one of the successful commercial CDNs, which hosts
some very popular websites, like Yahoo and Monster.com, and websites of many
big companies, like IBM and FedEx. \mycomment{However, Based on its' white
papers and some
papers published, we describe the conjecture on how Akamai works
as an representative instance of commercial CDNs,
although the details are not publicly
available as a business proprietary.}
By July 2002, Akamai has deployed more than 9,700 replica servers across 56
countries. These replica servers store the replication of the content for the
websites Akamai hosts. To achieve network diversity and proximity to users,
these replica servers sit in data centers and Points of Presence (POPs) of major
Internet and communication carriers. Akamai also built its own DNS network.
This Akamai DNS network
ensures fast delivery of the requested content by resolving the host name
of the URL for the requested content to the IP address of the Akamai replica
server that will deliver the desired content to the user most quickly.
Based on some white papers and published papers, we describe our conjectures on
two major processes involved in the operation of Akamai: one is how to direct
Internet traffic to the Akamai server network, another is how to direct
requests to the suitable Akamai replica servers.

\subsection{ARLs and Akamaizer}
Akamaizer is the tool that tags embedded Web objects for delivery via the Akamai
network, transforming ("akamaizing") their URLs into Akamai Resource Locators
(ARLs). ARLs contain a number of fields that aid in the content delivery
process. Their format is described in the following example.

A typical embedded object URL such as {\em http://www.foo.com/a.gif}
would be transformed into the following ARL:
\[http://a\overbrace{836}^\mathbf{\scriptscriptstyle Serial\,\#}.\overbrace{g.akamaitech.net}^\mathbf{\scriptscriptstyle Akamai\,Domain}/\overbrace{7}^\mathbf{\scriptscriptstyle Type}/\overbrace{836}^\mathbf{\scriptscriptstyle Serial\,\#}/\overbrace{123}^\mathbf{\scriptscriptstyle Provider\,Code}/\overbrace{e358f5db0045}^\mathbf{\scriptscriptstyle Object\,Data}/\overbrace{www.foo.com/a.gif}^\mathbf{\scriptscriptstyle absolute URL}\]
The {\em serial number} identifies a virtual "bucket" of content -- a group of
akamaized objects that will always be served from the same set of Akamai replica
servers. The {\em Akamai domain} ensures that requests for akamaized content
travel directly from the user to the Akamai network, completely avoiding the
object's origin site. The {\em type} field aids in interpreting an ARL. The
{\em provider code} uniquely identifies an Akamai customer.
The {\em object data} is used to guarantee object freshness. Depending on the
type in use, this field contains either the object's expiration time, or a 
string that uniquely identifies a particular version of the object, e.g., the
MD5 hash value of the object content. In the latter case, when the object is
modified, its object data field changes, so its ARL changes as well.
The last field {\em absolute URL} is used by Akamai replica servers to retrieve
the object from the content provider's origin site the first time the object
is requested.

\subsection{The Akamai DNS System}
All user requests for ARLs are directed to the Akamai network by the server
domain field (set to {\em g.akamai.net}) in each ARL. Then, the Akamai DNS
system chooses the Akamai replica server that will deliver the content to the
user most quickly and resolve the {\em *.g.akamai.net} server name using this
server's IP address. Unlike
the conventional DNS name resolution, this resolution relies not only on the
server name, but also on the source address of
the DNS query, current network condition and replica servers status.

The Akamai DNS system is implemented as a 2-level hierarchy of DNS servers:
by April 2000, 50 {\em high-level .akamai.net servers} (HLDNS) and 2000
{\em low-level .g.akamai.net servers} (LLDNS). Each HLDNS server is responsible
for directing each DNS query it receives to a LLDNS server that is close to
the requesting client. The LLDNS servers perform the final resolution of server
name to IP address, directing each client to the Akamai replica server
that is optimally located to serve the client's requests. As Akamai
continuously monitors network condition as well as the status of replica
servers, it can respond to network events within a matter of seconds.

When a browser makes a request for an ARL, whose server name is
{\em a9.g.akamai.net} for example, it first contacts its local DNS server,
asking it to resolve the server name. In the absence of a cached response, the
local DNS server resolves the server name by using iterative DNS queries. It
first contacts a {\em .net} root server, which responds with a list of Akamai
HLDNS servers. When the local DNS server contacts one of these HLDNS servers,
it receives
a list of LLDNS servers that are close to it. It then contacts one of the LLDNS
servers, which returns the IP address of the optimal replica server for this
request. Eventually, the local DNS server returns this IP address to the
requesting browser, which then fetches the content from that server.

The Akamai DNS system enables caching of DNS responses, just as in conventional
DNS name resolution, so as to avoid every request incurring the delay of three
levels of DNS queries. The Time-to-Live (TTL)
of the DNS responses are set in such a way
as to balance the benefits of caching with the chief goal of the Akamai DNS
system: keeping the client-to-server mapping
up to date with current network condition as well as replica servers status. As
the responses obtained from the root {\em .net} servers do not vary with
network conditions, they have a TTL of two days. The responses returned by
HLDNS servers are based on a network map that is recomputed every 7-10 minutes,
so these responses have a TTL of 20 minutes. Since LLDNS servers generate
name resolution based on maps that are updated every 2-10 seconds, the TTL 
in the responses from LLDNS servers is set to 20 seconds.



\section{Iridium: A Fast Content Location Service for Large-Scale Peer-to-Peer Systems}

\subsection{Related Work}
The present routing protocols or services in peer-to-peer systems can be
roughly classified into three categories according to the number of nodes
performing the routing operation. 
\begin{itemize}
\item The first category relies on a central
node for servicing all the routing requests in the system. Napster
\cite{Napster} and Audiogalaxy \cite{Audiogalaxy}
are the instances. Obviously, this centralized
structure can not scale well for large system and is error prone. A more
significant demerit of this structure is that the system using this routing
structure can be easily censored. The suspension of Napster shows this defect.
\item Some peer-to-peer systems
choose a small set of ``superpeers" or ``supernodes" by consensus to
service the routing requests in the system rather than replying on a central
node for routing.  Kazaa/Fasttrack \cite{Kazaa},
WinMx \cite{WinMx} and edonkey2000 \cite{edonkey2000} take this
approach. Gnutella \cite{Gnutella} also
has a slightly structured network for routing requests
but delivers requests in an expanding search, which incurs enormous
amount of search traffic. Fast content location generally is not an explicit
concern in these systems.
\item Tapestry \cite{zhao01tapestry}, CAN \cite{ratnasamy01scalable},
Chord \cite{Stoica} and Pastry \cite{rowstron01pastry}
constitute the peer-to-peer system in a completely
distributed structure where each node participates in the routing procedure.
They distribute the routing
information on each node. Each node maintains information only about $O(logN)$
\footnote{$N$ is the total number of nodes in the system.}
other nodes, and a request requires $O(logN)$ overlay hops to complete.
These systems
achieve the excellent scalability at the cost of large lookup latency. This
issue becomes worse in practice as each overlay hop
could take very long time, especially when
the physical path between the source and the sink of a hop is congested.
Brocade \cite{zhao02brocade} improves over Tapestry on this issue
by shortcutting the lookup messages through some
selected ``supernodes" which have high bandwidth and fast access to
the wide-area
network. However, it will still take $O(logN)$ overlay hops to finish a request.
\end{itemize}
To address the large lookup latency issue of Tapestry etc. have,
the proposed system in Iridium constructs
a routing fabric consisting of a set of ``supernodes".
The system is divided into multiple partitions each consisting of a supernode
and a set of regular nodes. Using consistent hashing \cite{karger97consistent},
the routing fabric takes constant time to find out which partition
stores the key being requested and service the request.
In the meanwhile, the routing system in Iridium scales well
as the routing operations are conducted in a distributed manner.

\subsection{System Structure}
Essentially, Iridium provides distributed computation of a hash function
mapping keys to nodes responsible for them within constant time. It uses
consistent hashing \cite{karger97consistent},
which has several good properties. With high
probability the hash function balances load. Also with high probability,
when an $N^{th}$ node joins (or leaves) the network, only an $O(1/N)$ fraction
of the keys are moved to a different location.

The consistent hash function assigns each node and key an $m$-bit $identifier$
using a base hash function such as SHA-1 \cite{sha}.
A node's identifier is chosen
by hashing the node's IP address, while a key identifier is calculated by
hashing the key. We will use the term ``key" to refer to
both the original key and
its identifier. Similarly, the term ``node" will refer to both the node and its
identifier under the hash function.

There are two types of nodes in Iridium: {\em supernode} and {\em regular node}.
\begin{itemize}
\item Regular nodes store the keys.
The keys are assigned by using the consistent
hashing as follows: All the identifiers are ordered in an
{\em identifier circle} modulo $2^m$. Key $k$ is assigned to the first regular
node whose identifier is equal to or follows $k$ in the identifier space. This
regular node is called the {\em successor node} of key $k$.
\item Supernodes store the node identifiers assigned by consistent hashing.
A node $n$ is assigned to the first supernode whose identifier is equal to
or follows $n$ in the supernode identifier space. This supernode is called the
{\em associated supernode} of node $n$. Correspondingly, we call the regular
nodes that are associated with supernode $s$ the {\em bound set} of $s$, denoted
as $b(s)$. A supernode performs the functionality of regular nodes as well.
\end{itemize}
To enhance the reliability, a key $k$ is stored in its successor node
and the $p-1$ regular nodes that follow key $k$'s successor node immediately.
These $p$ regular nodes together we call them the key $k$'s
{\em successor node set}. Similarly,
a node $n$ is assigned to its associated supernode and the
$q-1$ supernodes that follow node $n$'s associated supernode immediately.
We name these $q$ supernodes the node $n$'s {\em associated supernode set}.
Here $p$ and $q$ are tunable system parameters.
Supposing the total number of nodes in the system is $N$, we
denote the number of supernodes as $f(N)$ and
the average number of nodes each supernode
stores as $B(N)$, where $B(N)$ is equal to $q * N / f(N)$.

\myfig{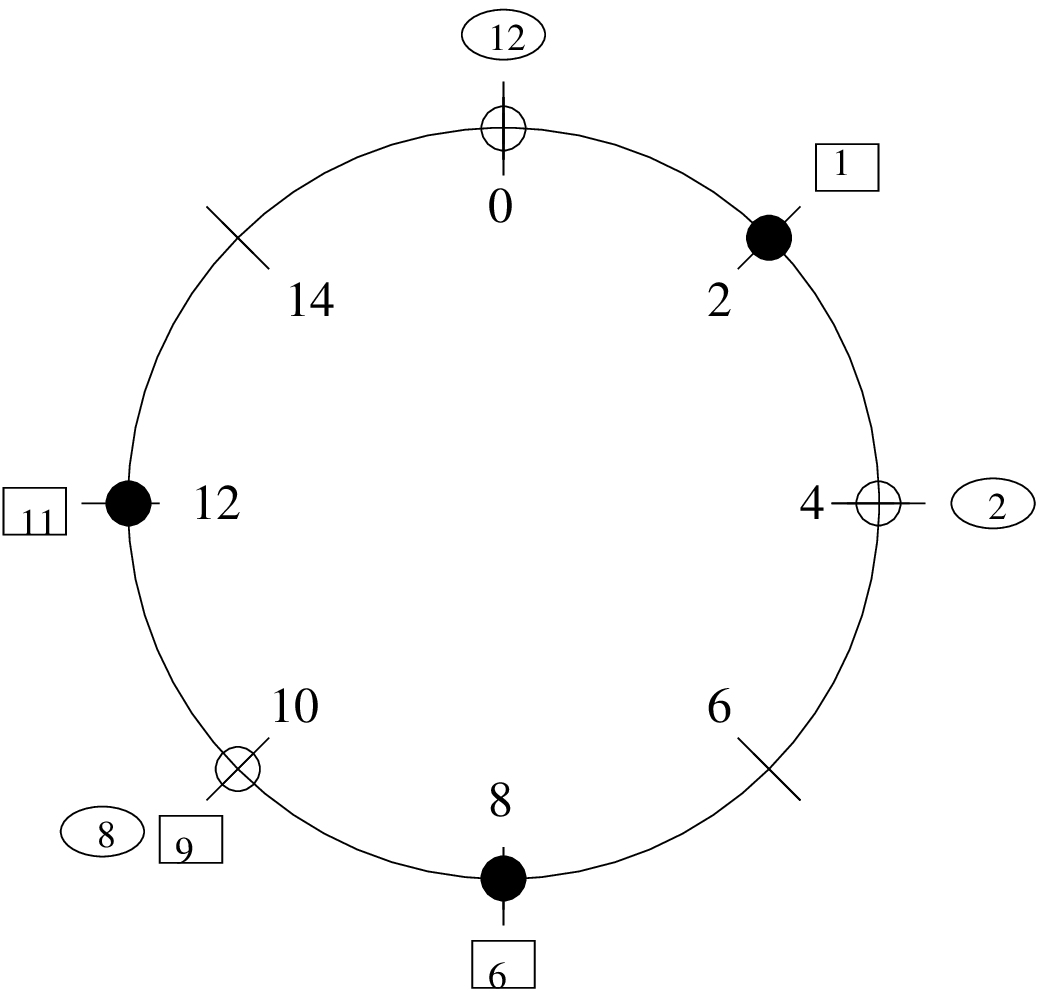}{1.9 in}
{An identifier circle consisting of the six nodes $0$, $2$,
$4$, $8$, $10$ and $11$. In them, node $2$, $8$ and $12$ are regular nodes and
node $0$, $4$ and $10$ are supernodes. Regular node $12$ is assigned to
supernode $0$, $2$ to $4$ and $8$ to $10$. Key $1$ is assigned to node $2$,
key $6$ to node $8$, key $9$ to node $10$ and key $11$ to node $12$.}

When a node $i$ is looking for a key $k$, it first randomly selects one
supernode, say $I$, from its associated supernode set to send a query to
asking for the location of key $k$. Supernode $I$ checks which supernode is
the clockwise closest supernode to key $k$ in the identifier space and
forwards the query to that supernode, say $J$.
Supernode $J$ then looks up which regular nodes in its bound set are holding
$k$ and
randomly chooses one, say $j$,
to deliver the query to. After finding the desired
key, the result will be sent back from $j$ to $J$, then $I$ and eventually
$i$. In total, each key lookup involves only constant number of hops.
For example, in Figure \ref{iridium:fig}, when node $2$ looks up for key $6$,
it first sends a query
to its associated supernode $4$. Supernode $4$ then sends this query to
supernode $10$, which is the clockwise closest supernode to key $6$ in the
identifier space. Supernode $10$ furthers 
the query to regular node $8$, which is the successor node of key $6$ and
stores the key $6$. Eventually, node $8$
sends the result through node $10$, $4$ back to the requesting node $2$.

\subsection{System Maintenance}

\subsubsection{Node Joins}
When a node wants to join the peer-to-peer system,
it needs to find out its associated
supernode first. It does this by flooding or broadcasting request looking for
a supernode to its topological neighbors. Once it gets the location of a
supernode, it queries the supernode for its associated supernode set.
After that,
it registers itself to its associated supernode set
by sending information to these supernodes including its identifier and address.

\subsubsection{Supernode Selection}
When the number of nodes in the system exceeds certain threshold or the routing
core consisting of the supernodes is overloaded,
we need to select new supernodes out of the regular nodes
to shed the workload.

First we determine the supernode from whose bound set to
choose a supernode candidate. The supernode could be the most loaded one or
the one having the largest bound set.
We then select a supernode candidate from the bound set.
We account the computation resources, including memory and CPU, and the
average up time when choosing a candidate. The time complexity of selecting
candidate is $B(N)$ if we search the bound set linearly
as each supernode will hold around $B(N)$
regular nodes. It could be optimized to constant time if each
supernode keeps its bound set sorted.
After a new supernode is selected, its previous associated supernode migrates
the regular nodes whose closest successor supernode becomes the
new supernode to the newly selected supernode.
\mycomment{This process could be done by sending one message in constant time.}
The next operation is to propagate the existence of this new supernode to other
supernodes. The simplest way to do this is to do broadcasting, which requires 
$f(N)$ messages for each supernode birth. This work could be optimized by
using lazy update. If a supernode needs to lookup some keys which are held by
certain regular nodes which
are in the bound set of the new supernode, it queries the old associated
supernode of the new supernode first. Besides forwarding the query to the
new supernode, the old associated supernode of the new supernode will sends
back response including the information about the new supernode to the
requesting supernode so that the information about the new supernode is
delivered to the interested supernodes gradually.

\subsubsection{Node Leaves}
Iridium handles the node leaving or death for supernodes and regular nodes in
different ways.
Normally, a regular node updates its identifier and address stored in its
associated supernode set either by explicitly sending message to its
associated supernode set periodically or attaching these information into the
query messages. The supernode simply
cache these information constructing the bound set. When a regular
node leaves, it notifies its associated supernode set its exit
so that the supernodes
can update their bound set. It also needs to migrate the keys
it stores to its successor node. When a regular node dies, its associated
supernode set will detect its death
by the timeout of the corresponding regular node information.
\mycomment{However, the keys stored on the dead node may get lost and we presume
the application utilizing the routing service has some mechanism to handle
this exception.}

It is more sophisticated to handle the leaving of supernodes as more information
need to be migrated or updated when a supernode leaves or dies compared to what we do in case of regular node leaving. Basically, when a supernode leaves,
it will
migrate its bound set to its living successor supernode. Also, it has to
tell the regular nodes to add one more successor
supernode to their associated supernode set to keep the size of the set
constant. To avoid the possible high instant traffic due to this operation,
each regular
node in the bound set of the leaving supernode waits for a random short period
before executing this operation.
\mycomment{The second operation will incur high instant traffic.
We alleviate
this problem by letting each regular node associate itself with a set of
successor
supernodes instead of just its closest successor supernode. We call this set
the regular node's {\em associated supernode set} and its closest successor
supernode its {\em private supernode}. This implies that the index
information of each regular node is replicated in several supernodes. This
replication will make the service more fault-tolerant, which we will discuss
further in the next subsection.}

When a supernode dies, other nodes could not notice
this event right at that moment. One straightforward approach to help the living
nodes detect the death of a supernode is to let each supernode broadcasts its
own identifier and address to other nodes periodically.
Obviously, this will incur too much traffic over the
Internet. For this reason, we take a lazy update approach to help detect the
death of a supernode but avoid the unnecessary message transmission.
This approach works as follows. After a supernode dies,
as the living supernodes do not know about its death,
they will forward the requests
for the key held by the bound set of the dead supernode to the dead supernode
as if it was alive. These requests will eventually timeout and as a result,
the living
supernodes forwarding the requests will find out that one supernode is down and
update their supernode information
accordingly. The same approach can be used by the regular nodes in the bound set
of the dead supernode to update their own associated supernode sets.
They send requests to the dead supernode as usual but
detect the death of their associated supernode by timeout.
Once they are aware of the death of their previous associated supernode,
they add one more successor supernode to their
associated supernode sets.
\mycomment{Once they are aware
of the death, they select the closest living successor supernode from their
associated supernode set as the new private supernode.} In this approach,
broadcasting the information of supernodes is avoided and also we do not need to
send the information about a supernode to the nodes that are not interested.

\subsection{Some Issues}

\paragraph {Reliability} 
When a regular node dies, the keys held by this node will lose.
Similarly, if a supernode dies, we lose information about the bound set of
this supernode. As a result of these information losing, the query will
be disrupted.
To guarantee the success of queries, we replicate information needed for
forwarding queries. In concrete, each key is replicated in
$p$ regular nodes. The queries for a key are almost equally shared by the
regular nodes holding the key. Similarly,
each regular node is associated with $q$ supernodes and anyone of these
supernodes has the same probability to
forward the query to or from the regular node.
To determine which values $p$ and $q$ should be so as to 
achieving an appropriate trade-off between reliability and scalability,
a simulation study is necessary.

\mycomment{
We think that the lost of regular node index is more serious than the lost of
content keys. The latter case could be handled by the application running on the
overlay. In order to reduce the probability of occurrence of the former case, we
should set some criteria, like reliability or average up time of the nodes,
when selecting supernodes. Supposing the probability of supernode failure is
$1/10$, which is a reasonable assumption given the supernode selection
procedure,
the probability of query failure due to supernode death is $10^{-q}$. When we
set $q$ to $8$, the probability of query being successful is $99.999999\%$. This
is satisfactory in practice. }

\mycomment{
As a query can only fail when either all the copies of the required content key
are gone or all the supernodes holding the index of the successor node list of
the required key are down. Taking the bigger of $p$ and $q$ as $r$, if the
probability of failure of one node is $1/2$, the probability of one query
failure is $2^{-r}$. When $r$ is $10$, which is a reasonable value,
the probability of query being success is $1023/1024$, which is bigger than
$99.9\%$. 
}

\mycomment{
\subsubsection{Recovery from Failure}
We think that the failure is inevitable even we replicate the information
necessary for routing to reasonably large extent. In case of node failure,
one question we need to answer to ensure a query would go through eventually is:
can we reconstruct the information when all the replication of one piece of
information are gone ?

The answer is yes.}

\paragraph {Scalability}
Unlike Tapestry etc., where each node participates in routing operation,
in Iridium, the routing work is taken by a limited set of supernodes.
As a result, Iridium may not scale as well as Tapestry etc. do.
One possible reason for this is the space each supernode has is limited.
However, to
be a first approximation, supposing $N$ is $10^7$, $f(N)$ chosen as
$N^{1/2}$, $q$ as $10$ and
each node information needs 12 bytes (6 bytes for IP address and 6 for
identifier), each supernode needs only about 760 KB to store the whole
bound set as well as the information about all the supernodes.
This amount of resource is
negligible in practice. Another issue which could possibly prevent
Iridium from scaling 
in practice is the amount of extra traffic supernodes may incur,
which includes the messages for routing and for maintenance of the system. To
answer this question, we need to perform extensive simulation study
under different functions of $f(N)$. On the other hand, regular routing load
is reduced heavily helping scalability.

\subsection{Summary}
In summary, Iridium is designed to provide fast content location service for
large-scale
peer-to-peer systems. As the design shows, the routing work is taken by a
set of supernodes leveraging consistent hashing and
a lookup can be done within constant time. We need to perform
extensive simulation experiments to study the scalability and reliability of
Iridium under various setups and scenarios.

\section{Summary}

With the increasing use of the Internet for content distribution,
caching and replication techniques are receiving more and more attention.
These are effective approaches to alleviate 
congestion on the Internet and to make the Internet more responsive. With many
unique features, CDN renders appreciable benefits to the content providers,
e.g., the popular web sites. In this report,
we studied the core challenging issues involved in
designing and building efficient CDNs, especially the content distribution
and request routing problems. A scheme for fast service location in peer-to-peer
systems is proposed.

Nowadays, new forms of Internet content and services, 
such as video-on-demand, which requires intensive bandwidth and predictable 
data transmission delay, are emerging. Meanwhile, the number
of content providers turning to CDNs to better service their customers is
growing rapidly. These two facts open several new issues to the 
design and architecture of CDNs in the future,
such as support for streaming content or real-time events,
scalability, built-in security mechanism, and so forth.
The research
on these issues are still undergoing and we believe that the successful
solutions to these problems will make the next-generation CDNs fly.

%


\bibliographystyle{alpha}
\bibliography{rpe}

\end{document}